\documentclass[twocolumn,superscriptaddress]{revtex4-2}
\usepackage{xcolor}
\usepackage{graphicx}
\usepackage{braket}
\usepackage{amsmath}
\usepackage{amssymb}
\usepackage{hyperref}
\usepackage[caption=false]{subfig}
\usepackage{tikz}
\usetikzlibrary{positioning}
\usetikzlibrary{shapes}
\definecolor{spinup}{rgb}{.9, .9, .0}
\definecolor{spindown}{rgb}{.0, 0.9, 0.9}
\newlength{\tikzscale}
\renewcommand{\vec}[1]{\boldsymbol{\mathbf{#1}}}
\newcommand{\CCO}{CaCuO\textsubscript{2}}

\begin{document}

\title{Capturing spin fluctuations in \CCO: \textit{Ab initio} QMC calculations with multi-determinant wave functions} 
\author{Chun Yu Chow}
\affiliation{Department of Physics, Institute for Condensed Matter Theory, University of Illinois at Urbana-Champaign, Urbana, Illinois 61801, USA} 
\author{William A. Wheeler}
\affiliation{Department of Materials Science and Engineering, University of Illinois at Urbana-Champaign, Urbana, Illinois 61801, USA} 
\author{Lucas K. Wagner}
\affiliation{Department of Physics, Institute for Condensed Matter Theory, University of Illinois at Urbana-Champaign, Urbana, Illinois 61801, USA} 
\begin{abstract}
    We present an advanced \textit{ab initio} quantum Monte Carlo (QMC) calculation of the ground state of undoped \CCO. 
We extend the traditional single-determinant Slater-Jastrow approach to include multi-determinant wave functions, inhomogeneous Jastrow factors, and orbital optimization. 
Our results demonstrate not only an improvement in the variational bound of the ground state energy -- 2.3 eV per formula unit lower than previous state-of-the-art techniques -- but also confirm the presence of spin fluctuations in multi-determinant wave functions in a strongly correlated cuprate system, which is integral to understanding high-$T_c$ superconductivity. 
This is the first demonstration of capturing spin fluctuations in QMC wave functions on a cuprate, establishing the groundwork for new studies on doped cuprates in the superconducting state, where spin fluctuations require more accurate characterization.
\end{abstract}
\maketitle

\section{Introduction}

Since the discovery of the first high-temperature cuprate superconductor in 1986 \cite{bednorzPossibleHighTcSuperconductivity1986}, understanding the mechanism for high-$T_c$ superconductivity has remained a major challenge for decades.
In the absence of holes, the undoped parent cuprates are antiferromagnetic (AFM) insulators.
Superconductivity emerges within a certain range of hole-doping, but the underlying mechanism is still under active debate \cite{bonnAreHightemperatureSuperconductors2006, leeDopingMottInsulator2006}.
Starting from early models such as Anderson's RVB theory \cite{andersonResonatingValenceBond1987}, the Zhang-Rice singlet \cite{zhangEffectiveHamiltonianSuperconducting1988}, and the Emery-Reiter spin polaron \cite{emeryTheoryHighMathrm1987, emeryMechanismHightemperatureSuperconductivity1988}, various modifications and the corresponding mechanisms have been proposed \cite{zhangValidityTJModel1990, mottSpinpolaronTheoryHighT1990, martinElectronicLocalizationCuprates1996, pavariniBandStructureTrendHoleDoped2001, macridinPhysicsCupratesTwoband2005, hozoiRenormalizationQuasiparticleHopping2007, pattersonSmallPolaronsMagnetic2008, peetsXRayAbsorptionSpectra2009, lauHighSpinPolaronLightly2011}.
Despite the proliferation of effective models, none have been accepted as a complete description of these materials.

\textit{Ab initio} methods offer a powerful approach to understanding materials \cite{feldtInitioMethodsFirstRow2022,ishikawaReviewInitioApproaches2015,dreuwSingleReferenceInitioMethods2005} but are notoriously difficult to apply to the cuprates.
Cuprates have strong correlation both at short range (so-called dynamic correlation or sometimes screening), and long range (so-called static correlation).
Recent work on these systems \cite{cuiSystematicElectronicStructure2022, cuiInitioQuantumManybody2023} has employed quantum chemistry techniques such as density matrix embedding theory (DMET) and coupled cluster theory, but the calculations used very small basis sets (double-zeta) which do not describe the short-range electronic correlation accurately.
While quantum Monte Carlo (QMC) methods are more flexible in describing electronic correlation, standard variational Monte Carlo (VMC) approaches for the cuprates have relied on some wave function parameters being set by a less accurate theory, for example, orbitals obtained from DFT \cite{wagnerEffectElectronCorrelation2014, foyevtsovaInitioQuantumMonte2014}.
For this reason, previous \textit{ab initio} QMC calculations that characterized low-energy spin configurations \cite{wagnerGroundStateDoped2015} were restricted to spin-symmetry-broken wave functions.
Recently, however, large multi-determinant expansions have become practical for realistic models of solids \cite{shinSystematicImprovementQuantum2024}, although their quality for the case of the cuprates is not known.

While the variational theorem provides a metric (energy) to measure progress in wave functions, it is important to investigate other properties that are integral to the physics of a given system.
Spin fluctuations have long been known to be relevant to the physics of cuprate superconductors \cite{mottSpinpolaronTheoryHighT1990, yuMagneticResonanceModel2010, scalapinoCommonThreadPairing2012}.
An ideal QMC calculation of cuprates should be able to both (i) achieve a low-energy variational upper bound by including short-range correlations and (ii) reproduce the spin fluctuations that are known to exist in the antiferromagnet.
Short-range correlations explain most of the correlation energy in many materials \cite{wagnerTransitionMetalOxides2007}, it therefore establishes the basis for the longer-range correlations that determine physical properties like spin fluctuations.
Despite the difference in energetic importance, short-range and long-range correlations interact and are inseparable; therefore, it is important to capture both in the calculation.

In this manuscript, we investigate the use of multideterminant expansions combined with orbital optimization and advanced Jastrow correlation factors \cite{sorellaWeakBindingTwo2007} to simultaneously treat short- and long-range correlation in the ground state of undoped \CCO \cite{dicastroHighSuperconductivityInterface2015}.
We confirm that the short-range correlation described by the Jastrow factor affects the one-particle physics of the system. 
We also show that a spin-symmetry-broken wave function as used in previous studies does not contain spin fluctuations expected in analogy to the Heisenberg model, whereas multi-determinant wave functions do achieve similar spin fluctuation behavior. 
Based on this success, we believe that advanced multi-determinant Slater-Jastrow wave functions are a promising tool to study the behavior of cuprates under doping.

\section{Method}
\subsection{Description of the system}

We carried out the calculations in this paper based on the non-relativistic electronic Hamiltonian under the Born-Oppenheimer approximation with fixed nuclei,
\begin{equation}
\hat{H} = -\frac{1}{2}\sum_i \nabla_i^2 + \sum_{ij} \frac{1}{r_{ij}} - \sum_{iA} \frac{Z_A}{r_{iA}},
\end{equation}
where $i$ and $A$ are the indices for electrons and nuclei respectively, $r_{ij}$ and $r_{iA}$ are the electron-electron and electron-nucleus distances, and $Z_A$ are the atomic numbers.

We used the lattice parameters from Ref.~\cite{wagnerEffectElectronCorrelation2014}, where the Cu-Cu distance is 3.86 {\AA} along the CuO$_2$ plane and 3.2 {\AA} between the planes.
The space group of this crystal is Pmm2, which means all the angles between the lattice vectors are 90$^\circ$.

\subsection{Density functional theory}
Our calculations started with density functional theory (DFT) calculations performed using the PySCF package \cite{sunRecentDevelopmentsPySCF2020, sunPySCFPythonbasedSimulations2018, sunLibcintEfficientGeneral2015}. 
We performed the calculations at the $\Gamma$ point of a $2\times 2 \times 1$ \CCO\, supercell.
The exchange-correlation functional Perdew–Burke–Ernzerhof-0 (PBE0) was chosen over PBE because of its lower diffusion Monte Carlo (DMC) energy for this system \cite{wagnerEffectElectronCorrelation2014}.
We used correlation-consistent effective core potentials (ECPs) and the corresponding triple-zeta basis sets (cc-pVTZ) from pseudopotentiallibrary.org \cite{bennettNewGenerationEffective2017, annaberdiyevNewGenerationEffective2018, wangNewGenerationEffective2019}.
We conducted two types of DFT calculations: unrestricted Kohn-Sham (UKS) and restricted Kohn-Sham (RKS).
The UKS calculations were initialized with an antiferromagnetic ordering of the Cu $3d_{x^2-y^2}$ spins, and retained this ordering through the convergence of the calculation. 
The RKS calculations were initialized to the converged UKS density and were converged to orbitals with no spin polarization.

\subsection{Complete active space self-consistent field}
The RKS orbitals are inputs to complete active space self-consistent field (CASSCF) calculations.
The CASSCF calculations were carried out using the PySCF package.
The active space was chosen to be the RKS orbitals near the Fermi energy with strong Cu $3d_{x^2-y^2}$ character, leading to a four orbital, four electron active space.
A larger CASSCF calculation was performed for reference using the atomic valence active space approach of Ref.~\cite{sayfutyarovaAutomatedConstructionMolecular2017} from Cu $3d_{x^2-y^2}$ and O $2p$ orbitals.
The natural orbitals from this calculation matched our four-orbital active space, validating the smaller active space calculation.

\subsection{Quantum Monte Carlo calculations}
The QMC calculations in this paper were performed using the PyQMC package \cite{wheelerPyQMCAllPythonRealspace2023}, where the many-body wave functions are implemented to take the general form
\begin{equation}
\Psi (\vec{R}) = e^{J(\vec{R},\vec{\alpha}, \vec{g})} \sum_k c_k D_k(\vec{R},\vec{\beta}), \label{eq:MSJ wave function}
\end{equation}
where $\vec{R}$ is the positions of all electrons, while $\vec{\alpha}$, $\vec{g}$, $\vec{c}$ and $\vec{\beta}$ are two-body Jastrow, geminal Jastrow, determinant expansion, and orbital parameters, respectively.
Every $D_k = \det \{ \phi_i^k(\vec{r}_j) \}$ represents a Slater determinant formed from a different set of single-particle orbitals $\{\phi_i\}$.
When there is more than one determinant, it is called a multi-Slater-Jastrow (MSJ) wave function.
A single-Slater-Jastrow (SJ) wave function is a special case of Eq.~\eqref{eq:MSJ wave function} when there is only one determinant.
The determinant in the SJ wave functions is formed from the UKS orbitals.
We truncated the CASSCF wave function up to the largest 10 determinants by the coefficients that dominate the superposition, which constitute the Slater part of our initial MSJ wave functions.

Two kinds of Jastrow factor were used in our calculations: the two-body Jastrow 
\begin{equation}
J_2(\vec{R},\vec{\alpha}) = \sum_{i < j} u(r_{ij}, \vec{\alpha}),
\end{equation}
used in all wave functions; and the geminal Jastrow \cite{sorellaWeakBindingTwo2007} 
\begin{equation}
J_G (\vec{R},\vec{g}) = 
\sum_{i<j} \sum_{mn} g_{mn} \phi_{m}(\vec{r}_i) \phi_{n}(\vec{r}_j),
\label{eq:geminal}
\end{equation}
for which we compare wave functions with and without. 
The two-body Jastrow factor depends only on pair distances between electrons and enforces the cusp condition \cite{wagnerEnergeticsDipoleMoment2007}, while the geminal Jastrow factor also takes into account the atomic environment around electron pairs.
Atomic Gaussian-type orbitals (GTOs) were chosen as the basis for the geminal Jastrow.
In particular, only $s$ orbitals were used, and the number of orbitals was chosen to balance accuracy and computational cost.
We will abbreviate an SJ wave function with only a two-body Jastrow as SJ2 and one with both a two-body and a geminal Jastrow as SJG.
Analogously, we denote MSJ wave functions by MSJ2 and MSJG, for a total of four wave function \textit{ans\"atze} in our QMC calculations.

After constructing the many-body wave functions, the parameters $\vec{\alpha}$, $\vec{g}$, $\vec{c}$, and $\vec{\beta}$ were optimized with variational Monte Carlo (VMC).
Optimizations were done in two steps: (i) Jastrow optimization -- only the Jastrow parameters $\vec{\alpha}$ (and $\vec{g}$ if applicable) were optimized; (ii) orbital optimization -- all parameters were optimized, including orbital parameters $\vec{\beta}$, as well as determinant coefficients $\vec{c}$ for MSJ wave functions.
Orbital optimizations were performed in batches of one to eight orbitals at a time to keep the number of parameters manageable.

Each of the optimized trial functions was then used in a fixed-node diffusion Monte Carlo (DMC) calculation,
which projects the trial function onto the ground state, subject to the constraint of having the same nodal surface as the initial function.
The DMC energies were extrapolated to the zero time step limit using calculations with time steps of 0.01, 0.02, 0.03, and 0.04  Ha$^{-1}$.
The extrapolation was done using the weighted least squares method with inverse variances as the weights.
The nonlocal parts of the Hamiltonian (ECPs) were treated with the T-move approximation \cite{casulaLocalityApproximationStandard2006, andersonNonlocalPseudopotentialsTimestep2021}.
Better parametrized wave functions should improve the T-move approximation because of its dependence on the accuracy of the trial wave function.

\subsection{Heisenberg model}
\begin{figure}
	\includegraphics{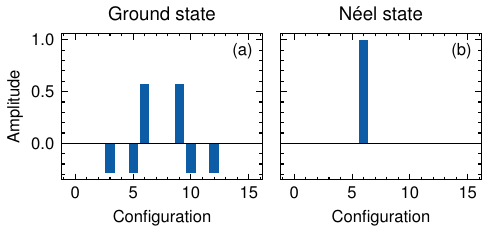}
    \caption{Superposition coefficients of (a) the Heisenberg ground state and (b) a Néel state. Spin configurations 6 and 9 are Néel states, while configurations 3, 5, 10, and 12 correspond to the stripe states.}
\label{fig:heisenberg_gs_neel}
\end{figure}

In addition to energy, we want to assess the spin properties of our trial wave functions.
Since the ground state of undoped cuprates is known to be antiferromagnetic (AFM), it is natural to compare the results of our \textit{ab initio} calculations against the ground state of an AFM Heisenberg model on a square lattice,
\begin{equation}
\hat{H} = J \sum_{\langle i, j\rangle} \hat{\vec{S}}_i \cdot \hat{\vec{S}}_j,
\end{equation}
where $\langle i, j \rangle$ are nearest-neighbor sites, $\hat{\vec{S}}_i$ is the spin operator on site $i$ (corresponding to a Cu $3d_{x^2-y^2}$ orbital), and $J > 0$ corresponds to AFM coupling. 
To compare against our \textit{ab initio} results, we considered the same $2 \times 2$ unit cell of the Heisenberg model with periodic boundary conditions.
The exact ground state of the model is a superposition of three different spin configurations: Néel states (checkerboard patterns -- the lowest energy classical spin configurations), horizontal stripe (h-stripe) states, and vertical stripe (v-stripe) states.
The superposition coefficients of the ground state and a Néel state in the basis of spin configurations are shown in Fig.~\ref{fig:heisenberg_gs_neel}, where the ground state is an equal superposition of the two Néel states plus negative contributions from the four possible stripe states.
These superpositions are known as spin fluctuations.

\subsection{Spin correlations} \label{subsec:spin correlations}

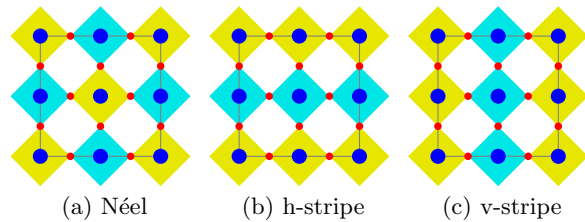
\begin{figure}
	\begin{center}
	\begin{tabular}{c c c} 
		\begin{tikzpicture}
[
Cu/.style={circle, fill=blue, inner sep=0mm, minimum size=0.5\tikzscale},
O/.style={circle, fill=red, inner sep=0mm, minimum size=0.25\tikzscale},
plaquette/.style={diamond, inner sep=0mm, minimum size=2\tikzscale},
plaqup/.style={plaquette, fill=spinup},
plaqdown/.style={plaquette, fill=spindown},
x=\tikzscale,y=\tikzscale,
]
\node (Cu0) at (0, 0) [plaqup]{};
\node (Cu1) at (0, 2) [plaqdown]{};
\node (Cu2) at (2, 2) [plaqup]{};
\node (Cu3) at (2, 0) [plaqdown]{};
\node (Cu04) at (0, 4) [plaqup]{};
\node (Cu44) at (4, 4) [plaqup]{};
\node (Cu40) at (4, 0) [plaqup]{};
\node (Cu42) at (4, 2) [plaqdown]{};
\node (Cu24) at (2, 4) [plaqdown]{};
\draw[gray] (0, 0) -- (4, 0) -- (4, 4) -- (0, 4) -- (0, 0);
\node at (1, 0) [O]{};
\node at (0, 1) [O]{};
\node at (2, 1) [O]{};
\node at (1, 2) [O]{};
\node at (3, 0) [O]{};
\node at (0, 3) [O]{};
\node at (2, 3) [O]{};
\node at (3, 2) [O]{};
\node at (1, 4) [O]{};
\node at (4, 1) [O]{};
\node at (3, 4) [O]{};
\node at (4, 3) [O]{};
\node at (Cu0.center) [Cu]{};
\node at (Cu1.center) [Cu]{};
\node at (Cu2.center) [Cu]{};
\node at (Cu3.center) [Cu]{};
\node at (Cu40.center) [Cu]{};
\node at (Cu44.center) [Cu]{};
\node at (Cu04.center) [Cu]{};
\node at (Cu42.center) [Cu]{};
\node at (Cu24.center) [Cu]{};

\end{tikzpicture} & \begin{tikzpicture}
[
Cu/.style={circle, fill=blue, inner sep=0mm, minimum size=0.5\tikzscale},
O/.style={circle, fill=red, inner sep=0mm, minimum size=0.25\tikzscale},
plaquette/.style={diamond, inner sep=0mm, minimum size=2\tikzscale},
plaqup/.style={plaquette, fill=spinup},
plaqdown/.style={plaquette, fill=spindown},
x=\tikzscale,y=\tikzscale,
]
\node (Cu0) at (0, 0) [plaqup]{};
\node (Cu1) at (0, 2) [plaqdown]{};
\node (Cu2) at (2, 2) [plaqdown]{};
\node (Cu3) at (2, 0) [plaqup]{};
\node (Cu04) at (0, 4) [plaqup]{};
\node (Cu44) at (4, 4) [plaqup]{};
\node (Cu40) at (4, 0) [plaqup]{};
\node (Cu42) at (4, 2) [plaqdown]{};
\node (Cu24) at (2, 4) [plaqup]{};
\draw[gray] (0, 0) -- (4, 0) -- (4, 4) -- (0, 4) -- (0, 0);
\node at (1, 0) [O]{};
\node at (0, 1) [O]{};
\node at (2, 1) [O]{};
\node at (1, 2) [O]{};
\node at (3, 0) [O]{};
\node at (0, 3) [O]{};
\node at (2, 3) [O]{};
\node at (3, 2) [O]{};
\node at (1, 4) [O]{};
\node at (4, 1) [O]{};
\node at (3, 4) [O]{};
\node at (4, 3) [O]{};
\node at (Cu0.center) [Cu]{};
\node at (Cu1.center) [Cu]{};
\node at (Cu2.center) [Cu]{};
\node at (Cu3.center) [Cu]{};
\node at (Cu40.center) [Cu]{};
\node at (Cu44.center) [Cu]{};
\node at (Cu04.center) [Cu]{};
\node at (Cu42.center) [Cu]{};
\node at (Cu24.center) [Cu]{};

\end{tikzpicture} & \begin{tikzpicture}
[
Cu/.style={circle, fill=blue, inner sep=0mm, minimum size=0.5\tikzscale},
O/.style={circle, fill=red, inner sep=0mm, minimum size=0.25\tikzscale},
plaquette/.style={diamond, inner sep=0mm, minimum size=2\tikzscale},
plaqup/.style={plaquette, fill=spinup},
plaqdown/.style={plaquette, fill=spindown},
x=\tikzscale,y=\tikzscale,
]
\node (Cu0) at (0, 0) [plaqup]{};
\node (Cu1) at (0, 2) [plaqup]{};
\node (Cu2) at (2, 2) [plaqdown]{};
\node (Cu3) at (2, 0) [plaqdown]{};
\node (Cu04) at (0, 4) [plaqup]{};
\node (Cu44) at (4, 4) [plaqup]{};
\node (Cu40) at (4, 0) [plaqup]{};
\node (Cu42) at (4, 2) [plaqup]{};
\node (Cu24) at (2, 4) [plaqdown]{};
\draw[gray] (0, 0) -- (4, 0) -- (4, 4) -- (0, 4) -- (0, 0);
\node at (1, 0) [O]{};
\node at (0, 1) [O]{};
\node at (2, 1) [O]{};
\node at (1, 2) [O]{};
\node at (3, 0) [O]{};
\node at (0, 3) [O]{};
\node at (2, 3) [O]{};
\node at (3, 2) [O]{};
\node at (1, 4) [O]{};
\node at (4, 1) [O]{};
\node at (3, 4) [O]{};
\node at (4, 3) [O]{};
\node at (Cu0.center) [Cu]{};
\node at (Cu1.center) [Cu]{};
\node at (Cu2.center) [Cu]{};
\node at (Cu3.center) [Cu]{};
\node at (Cu40.center) [Cu]{};
\node at (Cu44.center) [Cu]{};
\node at (Cu04.center) [Cu]{};
\node at (Cu42.center) [Cu]{};
\node at (Cu24.center) [Cu]{};

\end{tikzpicture} \\
		(a) Néel & (b) h-stripe & (c) v-stripe \\
	\end{tabular}
	\end{center}
	\caption{Spin correlations. (a) N\'eel correlation. (b)-(c) Stripe correlations. Each of the colored square plaquettes is a spin site where the spin population is counted. The positions of the atoms are represented by filled circles: Cu (blue) and O (red).}
\label{fig:patterns}
\end{figure}

To measure spin correlations in our \textit{ab initio} wave functions, we measured the spin population around the Cu atoms, which gives us an estimate of the magnetic moments that correspond to each site of the Heisenberg model.
The unit cell is divided into plaquettes surrounding the Cu atoms and plaquettes in the empty spaces, illustrated in Fig.~\ref{fig:patterns}. 
In each of the copper plaquettes, we measured the population of spin up and down electrons, whose sum and difference are the charge and spin populations, respectively.
Given an electronic configuration $\vec{R}$, we define a spin correlation as
\begin{equation}
\chi_{\text{p}} (\vec{R}) = \sum_{i} q_i^{\text{p}} \left( n_{i,\uparrow} - n_{i,\downarrow} \right),
\end{equation}
where $\text{p}$ is the type of correlation -- Néel, h-stripe, or v-stripe -- $i$ denotes the spin site, $q_i^{\text{p}} = \pm 1$ gives the signs according to the pattern $\text{p}$, and $n_{i,\uparrow / \downarrow}$ count the number of spin up or down electrons at spin site $i$ in $\vec{R}$.
The signs of the plaquettes $q_i^{\text{p}}$ correspond to the colors of the patterns in Fig.~\ref{fig:patterns}.

\section{Results} 

\subsection{Variational Monte Carlo}

\begin{figure}
    \includegraphics{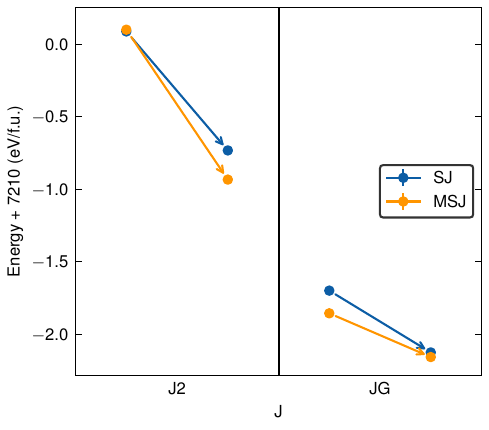}
    \caption{VMC energies from all optimized trial functions. The arrows indicate orbital optimization (as well as determinant coefficient optimization for MSJ wave functions) in addition to Jastrow optimization. Error bars show one standard deviation of statistical uncertainty.}
\label{fig:vmc}
\end{figure}

We have made three improvements to the previous state of the art for \textit{ab initio} cuprate calculations \cite{wagnerGroundStateDoped2015} (DFT-based symmetry-broken single determinant with two-body Jastrow factor): (i) use of multi-determinant wave functions, (ii) inclusion of a geminal Jastrow factor, and (iii) orbital optimization on the wave functions.
Comparing the MSJG wave function with orbitals optimized to the SJ2 wave function in Fig.~\ref{fig:vmc}, we see that the improved wave functions are up to 2.3 eV/f.u. (formula unit) lower in energy.
In order of energetic importance, including a geminal Jastrow factor makes the most significant impact, followed by orbital optimization, then the use of multi-determinant wave functions.
However, the multi-determinant wave functions are most important for obtaining accurate spin properties.

The addition of the geminal Jastrow factor to the two-body Jastrow factor results in significant improvements to all of the SJ and MSJ wave functions with and without orbital optimization (Fig.~\ref{fig:vmc}). 
The energy is reduced by  1.8 eV/f.u. without orbital optimization and by 1.4 eV/f.u. with orbital optimization, for both SJ and MSJ wave functions.
Orbital optimization decreases wave function energies by up to 1.0 eV/f.u. for the two-body Jastrow (SJ2, MSJ2) and 0.43 eV/f.u. for the geminal Jastrow (SJG, MSJG)  (Fig.~\ref{fig:vmc}).
Each of these improvements achieves an energy reduction of 5\% to 15\% of the VMC-DMC difference ($\sim 11$ eV/f.u.).

The orbital optimized SJG and MSJG wave functions have the lowest VMC energies than other functions by a significant margin.
The energy difference between these two is 0.13(6) eV per supercell, which is approximately equal to $J$ (0.14 eV) -- Heisenberg coupling constant -- of the system (computed with fixed-node DMC) \cite{wagnerEffectElectronCorrelation2014}.
The Heisenberg N\'eel state and ground state differ in energy by exactly $J$, showing a similarity in energy scale between the Heisenberg state pair and the \textit{ab initio} pair.
The symmetry-unbroken MSJ wave functions thus achieve lower energies by up to a few tenths of an eV/f.u. than the symmetry broken single determinant SJ wave functions.
This is non-trivial because the MSJ functions considered here are not a superset of the symmetry-broken SJ functions.

\begin{figure*}
    \includegraphics{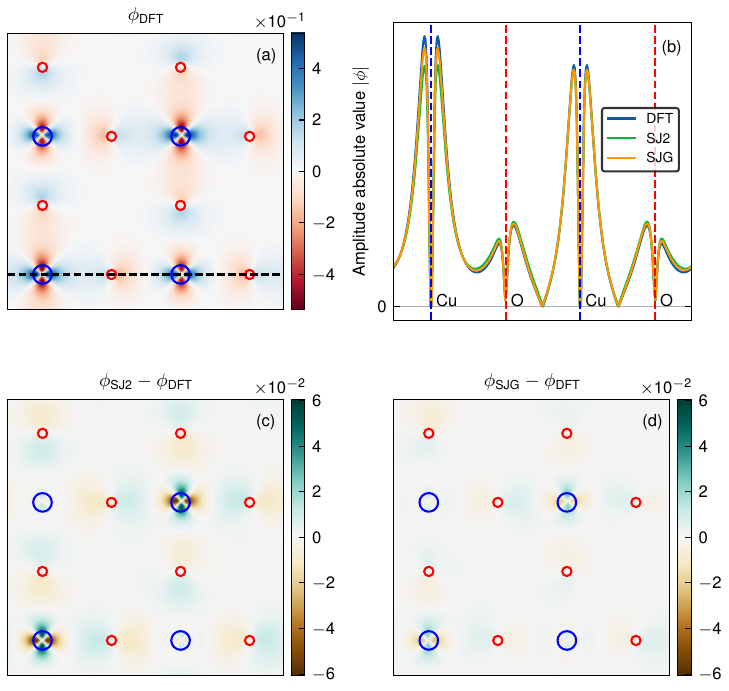}
    \caption{DFT orbital that is most changed by VMC orbital optimization. 
(a) The orbital amplitude over the 2$\times$2 calculation cell. Hollow circles indicate positions of Cu atoms (blue) and O atoms (red).
(b) Orbital along the black dashed line in (a) at different levels of optimization: $\phi_{\rm DFT}$,  $\phi_{\rm SJ2}$,  $\phi_{\rm SJG}$.
The vertical dashed lines indicate the position of Cu atoms (blue) and O atoms (red).
(c) Change in orbital from DFT after optimization with two-body Jastrow. The O lobes are increased, and the $3d_{x^2-y^2}$ lobes are decreased on two of the Cu atoms.
(d) Change in orbital from DFT after optimization with geminal Jastrow. The changes are qualitatively similar to those in (c), but smaller in magnitude (closer to the original DFT orbital).
    }
\label{fig:orbitals}
\end{figure*}

Orbital optimization impacts one-body properties in addition to energy.
Optimizing orbitals affects the bonding between Cu and O atoms, which is apparent in the changes of the orbital shapes. 
The orbital that changes most during optimization of the SJ2 and SJG wave functions is shown in Fig.~\ref{fig:orbitals}(a).
The absolute orbital amplitude $|\phi|$ decreases most in two of the Cu $d$ orbitals and increases slightly in O $p$ orbitals after optimization, seen from Fig.~\ref{fig:orbitals}(b), where $|\phi|$ is plotted along the black dashed line in the heatmap.
Fig.~\ref{fig:orbitals}(c) and (d), which show the difference in orbital amplitudes over the calculation cell before and after optimization, provide another perspective of seeing these changes.
The orbital weight transfer from Cu $d$ orbitals to O $p$ orbitals during the optimization indicates a change in the covalent character in the Cu-O bonds.
The changes in the SJG orbital are qualitatively similar to those in the SJ2 orbital but with a smaller magnitude, so the SJG orbital is more similar to the inital DFT orbital.
This is evidence that the effects of short-range correlations (the choice of Jastrow factor) propagate to the long-range structure of the wave functions.

\subsection{Diffusion Monte Carlo}

\begin{figure}
    \includegraphics{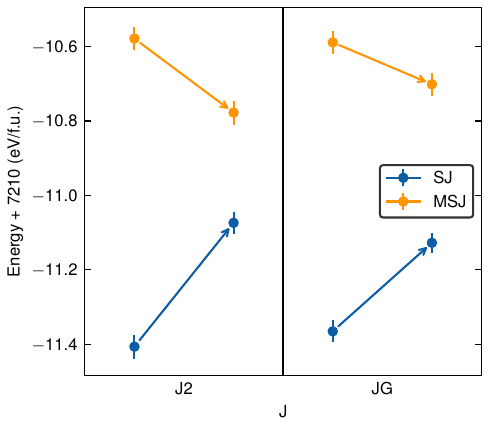}
    \caption{Fixed-node DMC energies from all optimized trial functions. The arrows indicate orbital optimization (as well as determinant coefficient optimization for MSJ wave functions) in addition to Jastrow optimization. Error bars show one standard deviation of statistical uncertainty.}
\label{fig:dmc}
\end{figure}

The effects of the geminal Jastrow, multiple determinants, and orbital optimization on the DMC energies are not as straightforward as they are in VMC.
First, we note that all of DMC energies in Fig.~\ref{fig:dmc} are close (within 0.9 eV/f.u.) compared to the spread of VMC energies (2.3 eV/f.u.) and compared to the VMC-DMC energy difference of $\sim 11$ eV/f.u.
The DMC energies of all trial functions show little difference within error bars in response to adding the geminal Jastrow factor, which suggests that the localization error is small.

Orbital optimization is clearly important for the MSJ functions considered here, because the starting orbitals from CASSCF were generated in the absence of a Jastrow factor, which affects the orbitals, as shown in Fig.~\ref{fig:orbitals}.
The DMC energies of SJ wave functions are higher after orbital optimization than before, because the orbitals were optimized in the presence of Jastrow correlations rather than the correlations present in DMC.
In fact, the PBE0 orbitals comprising the SJ functions were already optimized \emph{with respect to DMC correlations} over the hybrid functional mixing percentage \cite{wagnerGroundStateDoped2015}. Using Jastrow correlations is not as good and results in higher energy.
Regions near the nodes are low probability regions (by definition), and therefore contribute relatively little to the VMC energy, which prioritizes parameters that change higher probability parts of the wave function.
However, in DMC, the final energy is primarily determined by the nodal surface of the trial function.
The higher DMC energies indicate that lower VMC energies do not guarantee higher quality nodes.

The difference in DMC energy between MSJ and SJ wave functions is also due to the nodal surface.
Like with the SJ wave functions, Jastrow correlation is not as good as DMC correlation for optimizing  MSJ orbitals.
VMC optimization of the geminal Jastrow obtained only 15\% of the correlation energy between  the two-body Jastrow VMC and DMC energies.
Optimizing  orbitals of the more flexible MSJ function in VMC is not sufficient to overcome the DMC-optimized orbitals of the SJ function.
It is important to note that the SJ functions we used are symmetry-broken, and thus not a subset of the MSJ \textit{ansatz}, so achieving lower energy with the symmetry-preserving MSJ functions is not guaranteed even under perfect optimization.

In summary, it is crucial that the orbitals are optimized in the presence of electron correlations, whether through DFT functionals or directly in QMC.
Our initial MSJ orbitals were optimized in the absence of short-range correlations and thus were improved under VMC optimization.
As we show in the next section, MSJ wave functions correctly capture spin fluctuations, so orbital optimization is essential to advance beyond DFT spin-symmetry-broken wave functions.

\subsection{Spin fluctuations}

\begin{figure}
	\includegraphics{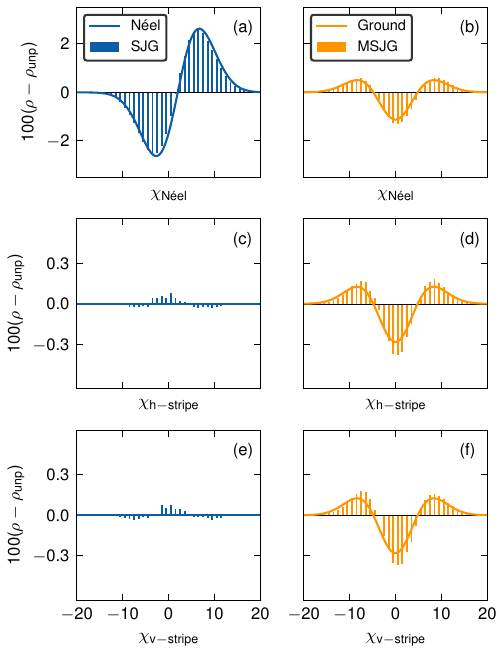}
	\caption{Distributions of spin correlations for the orbital optimized SJG and MSJG wave functions (histograms), and the Néel state and the ground state of the Heisenberg model (curves), with the spin-unpolarized distribution subtracted and scaled by $100$. 
	(a)-(b) N\'eel correlations.
	(c)-(f) Stripe correlations.
    The Heisenberg distributions are smoothed by Gaussians to fit the \textit{ab initio} distributions, using the same Gaussian width and height across all panels.
	The SJG wave function has similar spin correlations to the Néel state, while the MSJ has spin correlations similar to the true Heisenberg ground state.} 
\label{fig:distribution}
\end{figure}

In this section, we will show that the MSJ wave function achieves spin fluctuations very similar to the Heisenberg model ground state, while the SJ wave function shows spin correlations very similar to the symmetry-broken Néel state.

We compute the strength of three kinds of spin correlation -- N\'eel, h-stripe, and v-stripe -- for each of the wave functions.
In addition to the partially filled $3d_{x^2-y^2}$ orbitals represented in the Heisenberg model, 
the \textit{ab initio} calculations contain many electrons that do not contribute to the magnetic order.
In the continuum calculation, the positions of electrons fluctuate across the boundaries of the plaquette regions that define our spin populations, resulting in large fluctuations of the measured spin correlations.
To isolate the spin ordering present in each wave function $\Psi$, we compute the distributions $\rho$ of 
$\chi_{\text{N\'eel}}(\vec{R})$, $\chi_{\rm h-stripe}(\vec{R})$, and $\chi_{\rm v-stripe}(\vec{R})$ over configuration samples $\vec{R}\sim|\Psi(\vec{R})|^2$, and subtract off reference distributions computed from the spin-unpolarized RKS Slater determinant $\Psi_{\rm RKS}$.
These difference distributions (scaled by $100$) are shown as histograms in Fig.~\ref{fig:distribution}.
For clarity, only a single representative histogram is shown for each of SJ and MSJ wave function groups (orbital-optimized SJG and MSJG).
The other wave functions have nearly the same distributions.

For comparison, the corresponding distributions are computed for the Heisenberg model as well.
As with the \textit{ab initio} distributions, an unpolarized reference (delta function at zero) is subtracted off of each distribution.
The Heisenberg histograms are smoothed by Gaussians for comparison.
Fig.~\ref{fig:distribution} shows the \textit{ab initio} difference distributions (histograms) along with the smoothed Heisenberg differences (curves).
The width and height of the Gaussians are the same across all panels in Fig.~\ref{fig:distribution}, fit to minimize the combined root-mean-square error from the \textit{ab initio} difference histograms.

The symmetry-broken SJ wave function has the same spin character as the N\'eel state of the Heisenberg model.
The distribution of $\chi_\text{N\'eel}$ for the SJ function [Fig.~\ref{fig:distribution}(a)] is shifted to the right relative to the unpolarized distribution, reflecting broken spin symmetry and nonzero N\'eel correlation.
The Heisenberg N\'eel state matches the \textit{ab initio} quite closely, showing a similar shift to the right.
Stripe correlations (both $\chi_\text{h-stripe}$ and $\chi_\text{v-stripe}$) in the Heisenberg N\'eel state are exactly zero [equal to the reference; Fig.~\ref{fig:distribution}(c), (e)].
In the SJ wave function, stripe correlations differ little from the spin-unpolarized reference.
We interpret the small shift \emph{towards} $\chi_\text{stripe} = 0$ (from $\Psi_{\rm RKS}$ to $\Psi_{\rm SJ}$) as small signals of stripe order in the unpolarized $\Psi_{\rm RKS}$ that are suppressed in the magnetically ordered $\Psi_{\rm SJ}$.

The spin-symmetric MSJ wave function captures the spin character of the ground state of the Heisenberg model.
In contrast to the SJ distribution, the MSJ distribution is shifted away from zero symmetrically, with equal peaks at positive and negative $\chi_\text{N\'eel}$ [Fig.~\ref{fig:distribution} (b)].
The symmetric peaks about $\chi_\text{N\'eel}=0$ reflect the spin symmetry of the MSJ function,
and the shift away from zero indicates the presence of Néel correlation, despite having zero average $\chi_\text{N\'eel}$.
This pattern is matched by the Heisenberg ground state distribution.
Stripe correlations are similarly clear in the MSJ wave function [Fig.~\ref{fig:distribution}(d), (f)].
Both h- and v-stripe correlations  have the same distributions, consistent with the fourfold rotational symmetry of \CCO.
The Heisenberg ground state also exhibits stripe correlation, which matches the \textit{ab initio} distributions closely.

The SJ and MSJ \textit{ab initio} functions correspond closely to the N\'eel state and ground state of the Heisenberg model.
The Heisenberg spin correlation distributions qualitatively match the corresponding \textit{ab initio} distributions, seen clearly with the Gaussian smoothing.
In the MSJ wave function, stripe correlations are about six times smaller than the N\'eel correlation, while in the Heisenberg ground state, the peak heights between N\'eel and stripe differ by a factor of five, 
surprisingly consistent with our \textit{ab initio}  results.
The clear Néel and stripe correlations apparent in the MSJ wave functions -- and their near absence in the SJ wave functions -- show that the improved MSJ wave functions are effective at capturing the spin fluctuations corresponding to the $2 \times 2$ Heisenberg model.

\section{Conclusion} 

In summary, we have shown that spin fluctuations in the cuprates can be described at the \textit{ab initio} level for undoped \CCO\ by using small multi-reference wave functions.
In doing so, we obtained the best quality (in the sense of lowest energy) ground state wave functions of undoped \CCO\ to date. 
We found that short-range dynamic correlations play a critical role in the physics of the cuprates, and if ignored, result in significant qualitative errors in the long-range physics of the material.

Relative to the previous state of the art -- a fixed, symmetry-broken Slater determinant and a homogeneous two-body Jastrow -- our improved wave functions achieved a lowering of the ground state energy by 2.3 eV/f.u.
This is a significant reduction, obtaining about 20\% of the energy difference from the previous best VMC to fixed-node DMC energies.
The correlation introduced by the Jastrow factor (sometimes roughly called dynamic or short-range correlation) affects Cu-O hybridization -- a one-body property -- seen from the dependence of orbitals on the quality of the Jastrow factor.
Orbital optimization improves the energy of the MSJ wave functions when used as a nodal surface for diffusion Monte Carlo, but a density functional theory optimized on DMC energies obtains lower energy than orbitals optimized using a Jastrow factor.

With a multi-Slater-Jastrow (MSJ), spin fluctuations in the ground state of the \textit{ab initio} system closely resemble the 2$\times$2 Heisenberg model.
We established a clear correspondence between the symmetry-broken Slater-Jastrow wave function and the N\'eel state, and similarly between the multi-Slater-Jastrow (MSJ) wave function and the Heisenberg ground state.
The symmetry-broken \textit{ansatz} typically used in these materials appears to be unable to describe spin fluctuations.
If spin fluctuations are required for the description of superconductivity in these materials, a multi-reference or equivalent approach is necessary.

The characterization of spin fluctuations in undoped \CCO\ establishes a foundation for future studies on doped systems where the interplay between spin fluctuations and superconductivity is not well understood.
We found that a relatively small Slater determinant expansions, when paired with a powerful Jastrow factor, can describe spin fluctuations. 
Given this result, MSJ wave functions are promising for the \textit{ab initio} study of the doped materials.

\section*{Acknowledgements}

We would like to thank Alexander Muñoz for helpful discussions.
This work is supported by the U.S. National Science Foundation via Award No. 1931258 (W.A.W.) and by a grant from the Simons Foundation as part of the Simons Collaboration on the many-electron problem (C.Y.C, L.K.W.).
This work used computational resources of the Oak Ridge Leadership Computing Facility at the Oak Ridge National Laboratory, which is supported by the Office of Science of the U.S. Department of Energy under Contract No. DE-AC05-00OR22725,
as well as the Illinois Campus Cluster, a computing resource that is operated by the Illinois Campus Cluster Program (ICCP) in conjunction with the National Center for Supercomputing Applications (NCSA) and which is supported by funds from the University of Illinois at Urbana-Champaign.

\bibliography{ref.bib}

\begin{thebibliography}{41}%
\makeatletter
\providecommand \@ifxundefined [1]{%
 \@ifx{#1\undefined}
}%
\providecommand \@ifnum [1]{%
 \ifnum #1\expandafter \@firstoftwo
 \else \expandafter \@secondoftwo
 \fi
}%
\providecommand \@ifx [1]{%
 \ifx #1\expandafter \@firstoftwo
 \else \expandafter \@secondoftwo
 \fi
}%
\providecommand \natexlab [1]{#1}%
\providecommand \enquote  [1]{``#1''}%
\providecommand \bibnamefont  [1]{#1}%
\providecommand \bibfnamefont [1]{#1}%
\providecommand \citenamefont [1]{#1}%
\providecommand \href@noop [0]{\@secondoftwo}%
\providecommand \href [0]{\begingroup \@sanitize@url \@href}%
\providecommand \@href[1]{\@@startlink{#1}\@@href}%
\providecommand \@@href[1]{\endgroup#1\@@endlink}%
\providecommand \@sanitize@url [0]{\catcode `\\12\catcode `\$12\catcode
  `\&12\catcode `\#12\catcode `\^12\catcode `\_12\catcode `\%12\relax}%
\providecommand \@@startlink[1]{}%
\providecommand \@@endlink[0]{}%
\providecommand \url  [0]{\begingroup\@sanitize@url \@url }%
\providecommand \@url [1]{\endgroup\@href {#1}{\urlprefix }}%
\providecommand \urlprefix  [0]{URL }%
\providecommand \Eprint [0]{\href }%
\providecommand \doibase [0]{https://doi.org/}%
\providecommand \selectlanguage [0]{\@gobble}%
\providecommand \bibinfo  [0]{\@secondoftwo}%
\providecommand \bibfield  [0]{\@secondoftwo}%
\providecommand \translation [1]{[#1]}%
\providecommand \BibitemOpen [0]{}%
\providecommand \bibitemStop [0]{}%
\providecommand \bibitemNoStop [0]{.\EOS\space}%
\providecommand \EOS [0]{\spacefactor3000\relax}%
\providecommand \BibitemShut  [1]{\csname bibitem#1\endcsname}%
\let\auto@bib@innerbib\@empty
\bibitem [{\citenamefont {Bednorz}\ and\ \citenamefont
  {M{\"u}ller}(1986)}]{bednorzPossibleHighTcSuperconductivity1986}%
  \BibitemOpen
  \bibfield  {author} {\bibinfo {author} {\bibfnamefont {J.~G.}\ \bibnamefont
  {Bednorz}}\ and\ \bibinfo {author} {\bibfnamefont {K.~A.}\ \bibnamefont
  {M{\"u}ller}},\ }\bibfield  {title} {\bibinfo {title} {Possible {{highTc}}
  superconductivity in the {{Ba}}-{{La}}-{{Cu}}-{{O}} system},\ }\href
  {https://doi.org/10.1007/BF01303701} {\bibfield  {journal} {\bibinfo
  {journal} {Zeitschrift f{\"u}r Physik B Condensed Matter}\ }\textbf {\bibinfo
  {volume} {64}},\ \bibinfo {pages} {189} (\bibinfo {year} {1986})}\BibitemShut
  {NoStop}%
\bibitem [{\citenamefont
  {Bonn}(2006)}]{bonnAreHightemperatureSuperconductors2006}%
  \BibitemOpen
  \bibfield  {author} {\bibinfo {author} {\bibfnamefont {D.~A.}\ \bibnamefont
  {Bonn}},\ }\bibfield  {title} {\bibinfo {title} {Are high-temperature
  superconductors exotic?},\ }\href {https://doi.org/10.1038/nphys248}
  {\bibfield  {journal} {\bibinfo  {journal} {Nature Physics}\ }\textbf
  {\bibinfo {volume} {2}},\ \bibinfo {pages} {159} (\bibinfo {year}
  {2006})}\BibitemShut {NoStop}%
\bibitem [{\citenamefont {Lee}\ \emph {et~al.}(2006)\citenamefont {Lee},
  \citenamefont {Nagaosa},\ and\ \citenamefont
  {Wen}}]{leeDopingMottInsulator2006}%
  \BibitemOpen
  \bibfield  {author} {\bibinfo {author} {\bibfnamefont {P.~A.}\ \bibnamefont
  {Lee}}, \bibinfo {author} {\bibfnamefont {N.}~\bibnamefont {Nagaosa}},\ and\
  \bibinfo {author} {\bibfnamefont {X.-G.}\ \bibnamefont {Wen}},\ }\bibfield
  {title} {\bibinfo {title} {Doping a {{Mott}} insulator: {{Physics}} of
  high-temperature superconductivity},\ }\href
  {https://doi.org/10.1103/RevModPhys.78.17} {\bibfield  {journal} {\bibinfo
  {journal} {Reviews of Modern Physics}\ }\textbf {\bibinfo {volume} {78}},\
  \bibinfo {pages} {17} (\bibinfo {year} {2006})}\BibitemShut {NoStop}%
\bibitem [{\citenamefont {Anderson}(1987)}]{andersonResonatingValenceBond1987}%
  \BibitemOpen
  \bibfield  {author} {\bibinfo {author} {\bibfnamefont {P.~W.}\ \bibnamefont
  {Anderson}},\ }\bibfield  {title} {\bibinfo {title} {The {{Resonating Valence
  Bond State}} in {{La2CuO4}} and {{Superconductivity}}},\ }\href
  {https://doi.org/10.1126/science.235.4793.1196} {\bibfield  {journal}
  {\bibinfo  {journal} {Science}\ }\textbf {\bibinfo {volume} {235}},\ \bibinfo
  {pages} {1196} (\bibinfo {year} {1987})}\BibitemShut {NoStop}%
\bibitem [{\citenamefont {Zhang}\ and\ \citenamefont
  {Rice}(1988)}]{zhangEffectiveHamiltonianSuperconducting1988}%
  \BibitemOpen
  \bibfield  {author} {\bibinfo {author} {\bibfnamefont {F.~C.}\ \bibnamefont
  {Zhang}}\ and\ \bibinfo {author} {\bibfnamefont {T.~M.}\ \bibnamefont
  {Rice}},\ }\bibfield  {title} {\bibinfo {title} {Effective {{Hamiltonian}}
  for the superconducting {{Cu}} oxides},\ }\href
  {https://doi.org/10.1103/PhysRevB.37.3759} {\bibfield  {journal} {\bibinfo
  {journal} {Physical Review B}\ }\textbf {\bibinfo {volume} {37}},\ \bibinfo
  {pages} {3759} (\bibinfo {year} {1988})}\BibitemShut {NoStop}%
\bibitem [{\citenamefont {Emery}(1987)}]{emeryTheoryHighMathrm1987}%
  \BibitemOpen
  \bibfield  {author} {\bibinfo {author} {\bibfnamefont {V.~J.}\ \bibnamefont
  {Emery}},\ }\bibfield  {title} {\bibinfo {title} {Theory of
  high-{${\mathrm{T}}_{\mathrm{c}}$} superconductivity in oxides},\ }\href
  {https://doi.org/10.1103/PhysRevLett.58.2794} {\bibfield  {journal} {\bibinfo
   {journal} {Physical Review Letters}\ }\textbf {\bibinfo {volume} {58}},\
  \bibinfo {pages} {2794} (\bibinfo {year} {1987})}\BibitemShut {NoStop}%
\bibitem [{\citenamefont {Emery}\ and\ \citenamefont
  {Reiter}(1988)}]{emeryMechanismHightemperatureSuperconductivity1988}%
  \BibitemOpen
  \bibfield  {author} {\bibinfo {author} {\bibfnamefont {V.~J.}\ \bibnamefont
  {Emery}}\ and\ \bibinfo {author} {\bibfnamefont {G.}~\bibnamefont {Reiter}},\
  }\bibfield  {title} {\bibinfo {title} {Mechanism for high-temperature
  superconductivity},\ }\href {https://doi.org/10.1103/PhysRevB.38.4547}
  {\bibfield  {journal} {\bibinfo  {journal} {Physical Review B}\ }\textbf
  {\bibinfo {volume} {38}},\ \bibinfo {pages} {4547} (\bibinfo {year}
  {1988})}\BibitemShut {NoStop}%
\bibitem [{\citenamefont {Zhang}\ and\ \citenamefont
  {Rice}(1990)}]{zhangValidityTJModel1990}%
  \BibitemOpen
  \bibfield  {author} {\bibinfo {author} {\bibfnamefont {F.~C.}\ \bibnamefont
  {Zhang}}\ and\ \bibinfo {author} {\bibfnamefont {T.~M.}\ \bibnamefont
  {Rice}},\ }\bibfield  {title} {\bibinfo {title} {Validity of the t-{{J}}
  model},\ }\href {https://doi.org/10.1103/PhysRevB.41.7243} {\bibfield
  {journal} {\bibinfo  {journal} {Physical Review B}\ }\textbf {\bibinfo
  {volume} {41}},\ \bibinfo {pages} {7243} (\bibinfo {year}
  {1990})}\BibitemShut {NoStop}%
\bibitem [{\citenamefont {Mott}(1990)}]{mottSpinpolaronTheoryHighT1990}%
  \BibitemOpen
  \bibfield  {author} {\bibinfo {author} {\bibfnamefont {N.}~\bibnamefont
  {Mott}},\ }\bibfield  {title} {\bibinfo {title} {The spin-polaron theory of
  high-{{T}} c superconductivity},\ }\href
  {https://doi.org/10.1080/00018739000101471} {\bibfield  {journal} {\bibinfo
  {journal} {Advances in Physics}\ }\textbf {\bibinfo {volume} {39}},\ \bibinfo
  {pages} {55} (\bibinfo {year} {1990})}\BibitemShut {NoStop}%
\bibitem [{\citenamefont
  {Martin}(1996)}]{martinElectronicLocalizationCuprates1996}%
  \BibitemOpen
  \bibfield  {author} {\bibinfo {author} {\bibfnamefont {R.~L.}\ \bibnamefont
  {Martin}},\ }\bibfield  {title} {\bibinfo {title} {Electronic localization in
  the cuprates},\ }\href {https://doi.org/10.1103/PhysRevB.53.15501} {\bibfield
   {journal} {\bibinfo  {journal} {Physical Review B}\ }\textbf {\bibinfo
  {volume} {53}},\ \bibinfo {pages} {15501} (\bibinfo {year}
  {1996})}\BibitemShut {NoStop}%
\bibitem [{\citenamefont {Pavarini}\ \emph {et~al.}(2001)\citenamefont
  {Pavarini}, \citenamefont {Dasgupta}, \citenamefont {{Saha-Dasgupta}},
  \citenamefont {Jepsen},\ and\ \citenamefont
  {Andersen}}]{pavariniBandStructureTrendHoleDoped2001}%
  \BibitemOpen
  \bibfield  {author} {\bibinfo {author} {\bibfnamefont {E.}~\bibnamefont
  {Pavarini}}, \bibinfo {author} {\bibfnamefont {I.}~\bibnamefont {Dasgupta}},
  \bibinfo {author} {\bibfnamefont {T.}~\bibnamefont {{Saha-Dasgupta}}},
  \bibinfo {author} {\bibfnamefont {O.}~\bibnamefont {Jepsen}},\ and\ \bibinfo
  {author} {\bibfnamefont {O.~K.}\ \bibnamefont {Andersen}},\ }\bibfield
  {title} {\bibinfo {title} {Band-structure trend in hole-doped cuprates and
  correlation with {${\mathit{T}}_{\mathit{c}\mathrm{max}}$}},\ }\href
  {https://doi.org/10.1103/PhysRevLett.87.047003} {\bibfield  {journal}
  {\bibinfo  {journal} {Physical Review Letters}\ }\textbf {\bibinfo {volume}
  {87}},\ \bibinfo {pages} {047003} (\bibinfo {year} {2001})}\BibitemShut
  {NoStop}%
\bibitem [{\citenamefont {Macridin}\ \emph {et~al.}(2005)\citenamefont
  {Macridin}, \citenamefont {Jarrell}, \citenamefont {Maier},\ and\
  \citenamefont {Sawatzky}}]{macridinPhysicsCupratesTwoband2005}%
  \BibitemOpen
  \bibfield  {author} {\bibinfo {author} {\bibfnamefont {A.}~\bibnamefont
  {Macridin}}, \bibinfo {author} {\bibfnamefont {M.}~\bibnamefont {Jarrell}},
  \bibinfo {author} {\bibfnamefont {{\relax Th}.}~\bibnamefont {Maier}},\ and\
  \bibinfo {author} {\bibfnamefont {G.~A.}\ \bibnamefont {Sawatzky}},\
  }\bibfield  {title} {\bibinfo {title} {Physics of cuprates with the two-band
  {{Hubbard}} model: {{The}} validity of the one-band {{Hubbard}} model},\
  }\href {https://doi.org/10.1103/PhysRevB.71.134527} {\bibfield  {journal}
  {\bibinfo  {journal} {Physical Review B}\ }\textbf {\bibinfo {volume} {71}},\
  \bibinfo {pages} {134527} (\bibinfo {year} {2005})}\BibitemShut {NoStop}%
\bibitem [{\citenamefont {Hozoi}\ \emph {et~al.}(2007)\citenamefont {Hozoi},
  \citenamefont {Nishimoto},\ and\ \citenamefont {{de
  Graaf}}}]{hozoiRenormalizationQuasiparticleHopping2007}%
  \BibitemOpen
  \bibfield  {author} {\bibinfo {author} {\bibfnamefont {L.}~\bibnamefont
  {Hozoi}}, \bibinfo {author} {\bibfnamefont {S.}~\bibnamefont {Nishimoto}},\
  and\ \bibinfo {author} {\bibfnamefont {C.}~\bibnamefont {{de Graaf}}},\
  }\bibfield  {title} {\bibinfo {title} {Renormalization of quasiparticle
  hopping integrals by spin interactions in layered copper oxides},\ }\href
  {https://doi.org/10.1103/PhysRevB.75.174505} {\bibfield  {journal} {\bibinfo
  {journal} {Physical Review B}\ }\textbf {\bibinfo {volume} {75}},\ \bibinfo
  {pages} {174505} (\bibinfo {year} {2007})}\BibitemShut {NoStop}%
\bibitem [{\citenamefont
  {Patterson}(2008)}]{pattersonSmallPolaronsMagnetic2008}%
  \BibitemOpen
  \bibfield  {author} {\bibinfo {author} {\bibfnamefont {C.~H.}\ \bibnamefont
  {Patterson}},\ }\bibfield  {title} {\bibinfo {title} {Small polarons and
  magnetic antiphase boundaries in
  {${\mathrm{Ca}}_{2\ensuremath{-}x}{\mathrm{Na}}_{x}\mathrm{Cu}{\mathrm{O}}_{2}{\mathrm{Cl}}_{2}$}
  {$(x=0.06,0.12)$}: Origin of striped phases in cuprates},\ }\href
  {https://doi.org/10.1103/PhysRevB.77.094523} {\bibfield  {journal} {\bibinfo
  {journal} {Physical Review B}\ }\textbf {\bibinfo {volume} {77}},\ \bibinfo
  {pages} {094523} (\bibinfo {year} {2008})}\BibitemShut {NoStop}%
\bibitem [{\citenamefont {Peets}\ \emph {et~al.}(2009)\citenamefont {Peets},
  \citenamefont {Hawthorn}, \citenamefont {Shen}, \citenamefont {Kim},
  \citenamefont {Ellis}, \citenamefont {Zhang}, \citenamefont {Komiya},
  \citenamefont {Ando}, \citenamefont {Sawatzky}, \citenamefont {Liang},
  \citenamefont {Bonn},\ and\ \citenamefont
  {Hardy}}]{peetsXRayAbsorptionSpectra2009}%
  \BibitemOpen
  \bibfield  {author} {\bibinfo {author} {\bibfnamefont {D.~C.}\ \bibnamefont
  {Peets}}, \bibinfo {author} {\bibfnamefont {D.~G.}\ \bibnamefont {Hawthorn}},
  \bibinfo {author} {\bibfnamefont {K.~M.}\ \bibnamefont {Shen}}, \bibinfo
  {author} {\bibfnamefont {Y.-J.}\ \bibnamefont {Kim}}, \bibinfo {author}
  {\bibfnamefont {D.~S.}\ \bibnamefont {Ellis}}, \bibinfo {author}
  {\bibfnamefont {H.}~\bibnamefont {Zhang}}, \bibinfo {author} {\bibfnamefont
  {S.}~\bibnamefont {Komiya}}, \bibinfo {author} {\bibfnamefont
  {Y.}~\bibnamefont {Ando}}, \bibinfo {author} {\bibfnamefont {G.~A.}\
  \bibnamefont {Sawatzky}}, \bibinfo {author} {\bibfnamefont {R.}~\bibnamefont
  {Liang}}, \bibinfo {author} {\bibfnamefont {D.~A.}\ \bibnamefont {Bonn}},\
  and\ \bibinfo {author} {\bibfnamefont {W.~N.}\ \bibnamefont {Hardy}},\
  }\bibfield  {title} {\bibinfo {title} {X-{{Ray Absorption Spectra Reveal}}
  the {{Inapplicability}} of the {{Single-Band Hubbard Model}} to {{Overdoped
  Cuprate Superconductors}}},\ }\href
  {https://doi.org/10.1103/PhysRevLett.103.087402} {\bibfield  {journal}
  {\bibinfo  {journal} {Physical Review Letters}\ }\textbf {\bibinfo {volume}
  {103}},\ \bibinfo {pages} {087402} (\bibinfo {year} {2009})}\BibitemShut
  {NoStop}%
\bibitem [{\citenamefont {Lau}\ \emph {et~al.}(2011)\citenamefont {Lau},
  \citenamefont {Berciu},\ and\ \citenamefont
  {Sawatzky}}]{lauHighSpinPolaronLightly2011}%
  \BibitemOpen
  \bibfield  {author} {\bibinfo {author} {\bibfnamefont {B.}~\bibnamefont
  {Lau}}, \bibinfo {author} {\bibfnamefont {M.}~\bibnamefont {Berciu}},\ and\
  \bibinfo {author} {\bibfnamefont {G.~A.}\ \bibnamefont {Sawatzky}},\
  }\bibfield  {title} {\bibinfo {title} {High-spin polaron in lightly doped
  {${\mathrm{CuO}}_{2}$} planes},\ }\href
  {https://doi.org/10.1103/PhysRevLett.106.036401} {\bibfield  {journal}
  {\bibinfo  {journal} {Physical Review Letters}\ }\textbf {\bibinfo {volume}
  {106}},\ \bibinfo {pages} {036401} (\bibinfo {year} {2011})}\BibitemShut
  {NoStop}%
\bibitem [{\citenamefont {Feldt}\ and\ \citenamefont
  {Phung}(2022)}]{feldtInitioMethodsFirstRow2022}%
  \BibitemOpen
  \bibfield  {author} {\bibinfo {author} {\bibfnamefont {M.}~\bibnamefont
  {Feldt}}\ and\ \bibinfo {author} {\bibfnamefont {Q.~M.}\ \bibnamefont
  {Phung}},\ }\bibfield  {title} {\bibinfo {title} {Ab {{Initio Methods}} in
  {{First-Row Transition Metal Chemistry}}},\ }\href
  {https://doi.org/10.1002/ejic.202200014} {\bibfield  {journal} {\bibinfo
  {journal} {European Journal of Inorganic Chemistry}\ }\textbf {\bibinfo
  {volume} {2022}},\ \bibinfo {pages} {e202200014} (\bibinfo {year}
  {2022})}\BibitemShut {NoStop}%
\bibitem [{\citenamefont {Ishikawa}\ and\ \citenamefont
  {Sato}(2015)}]{ishikawaReviewInitioApproaches2015}%
  \BibitemOpen
  \bibfield  {author} {\bibinfo {author} {\bibfnamefont {K.~L.}\ \bibnamefont
  {Ishikawa}}\ and\ \bibinfo {author} {\bibfnamefont {T.}~\bibnamefont
  {Sato}},\ }\bibfield  {title} {\bibinfo {title} {A {{Review}} on {{Ab Initio
  Approaches}} for {{Multielectron Dynamics}}},\ }\href
  {https://doi.org/10.1109/JSTQE.2015.2438827} {\bibfield  {journal} {\bibinfo
  {journal} {IEEE Journal of Selected Topics in Quantum Electronics}\ }\textbf
  {\bibinfo {volume} {21}},\ \bibinfo {pages} {1} (\bibinfo {year}
  {2015})}\BibitemShut {NoStop}%
\bibitem [{\citenamefont {Dreuw}\ and\ \citenamefont
  {{Head-Gordon}}(2005)}]{dreuwSingleReferenceInitioMethods2005}%
  \BibitemOpen
  \bibfield  {author} {\bibinfo {author} {\bibfnamefont {A.}~\bibnamefont
  {Dreuw}}\ and\ \bibinfo {author} {\bibfnamefont {M.}~\bibnamefont
  {{Head-Gordon}}},\ }\bibfield  {title} {\bibinfo {title}
  {Single-{{Reference}} ab {{Initio Methods}} for the {{Calculation}} of
  {{Excited States}} of {{Large Molecules}}},\ }\href
  {https://doi.org/10.1021/cr0505627} {\bibfield  {journal} {\bibinfo
  {journal} {Chemical Reviews}\ }\textbf {\bibinfo {volume} {105}},\ \bibinfo
  {pages} {4009} (\bibinfo {year} {2005})}\BibitemShut {NoStop}%
\bibitem [{\citenamefont {Cui}\ \emph {et~al.}(2022)\citenamefont {Cui},
  \citenamefont {Zhai}, \citenamefont {Zhang},\ and\ \citenamefont
  {Chan}}]{cuiSystematicElectronicStructure2022}%
  \BibitemOpen
  \bibfield  {author} {\bibinfo {author} {\bibfnamefont {Z.-H.}\ \bibnamefont
  {Cui}}, \bibinfo {author} {\bibfnamefont {H.}~\bibnamefont {Zhai}}, \bibinfo
  {author} {\bibfnamefont {X.}~\bibnamefont {Zhang}},\ and\ \bibinfo {author}
  {\bibfnamefont {G.~K.-L.}\ \bibnamefont {Chan}},\ }\bibfield  {title}
  {\bibinfo {title} {Systematic electronic structure in the cuprate parent
  state from quantum many-body simulations},\ }\href
  {https://doi.org/10.1126/science.abm2295} {\bibfield  {journal} {\bibinfo
  {journal} {Science}\ }\textbf {\bibinfo {volume} {377}},\ \bibinfo {pages}
  {1192} (\bibinfo {year} {2022})}\BibitemShut {NoStop}%
\bibitem [{\citenamefont {Cui}\ \emph {et~al.}(2023)\citenamefont {Cui},
  \citenamefont {Yang}, \citenamefont {T{\"o}lle}, \citenamefont {Ye},
  \citenamefont {Zhai}, \citenamefont {Kim}, \citenamefont {Zhang},
  \citenamefont {Lin}, \citenamefont {Berkelbach},\ and\ \citenamefont
  {Chan}}]{cuiInitioQuantumManybody2023}%
  \BibitemOpen
  \bibfield  {author} {\bibinfo {author} {\bibfnamefont {Z.-H.}\ \bibnamefont
  {Cui}}, \bibinfo {author} {\bibfnamefont {J.}~\bibnamefont {Yang}}, \bibinfo
  {author} {\bibfnamefont {J.}~\bibnamefont {T{\"o}lle}}, \bibinfo {author}
  {\bibfnamefont {H.-Z.}\ \bibnamefont {Ye}}, \bibinfo {author} {\bibfnamefont
  {H.}~\bibnamefont {Zhai}}, \bibinfo {author} {\bibfnamefont {R.}~\bibnamefont
  {Kim}}, \bibinfo {author} {\bibfnamefont {X.}~\bibnamefont {Zhang}}, \bibinfo
  {author} {\bibfnamefont {L.}~\bibnamefont {Lin}}, \bibinfo {author}
  {\bibfnamefont {T.~C.}\ \bibnamefont {Berkelbach}},\ and\ \bibinfo {author}
  {\bibfnamefont {G.~K.-L.}\ \bibnamefont {Chan}},\ }\href
  {https://doi.org/10.48550/arXiv.2306.16561} {\bibinfo {title} {Ab initio
  quantum many-body description of superconducting trends in the cuprates}}
  (\bibinfo {year} {2023}),\ \Eprint {https://arxiv.org/abs/2306.16561}
  {arXiv:2306.16561} \BibitemShut {NoStop}%
\bibitem [{\citenamefont {Wagner}\ and\ \citenamefont
  {Abbamonte}(2014)}]{wagnerEffectElectronCorrelation2014}%
  \BibitemOpen
  \bibfield  {author} {\bibinfo {author} {\bibfnamefont {L.~K.}\ \bibnamefont
  {Wagner}}\ and\ \bibinfo {author} {\bibfnamefont {P.}~\bibnamefont
  {Abbamonte}},\ }\bibfield  {title} {\bibinfo {title} {Effect of electron
  correlation on the electronic structure and spin-lattice coupling of high-
  {{T}} c cuprates: {{Quantum Monte Carlo}} calculations},\ }\href
  {https://doi.org/10.1103/PhysRevB.90.125129} {\bibfield  {journal} {\bibinfo
  {journal} {Physical Review B}\ }\textbf {\bibinfo {volume} {90}},\ \bibinfo
  {pages} {125129} (\bibinfo {year} {2014})}\BibitemShut {NoStop}%
\bibitem [{\citenamefont {Foyevtsova}\ \emph {et~al.}(2014)\citenamefont
  {Foyevtsova}, \citenamefont {Krogel}, \citenamefont {Kim}, \citenamefont
  {Kent}, \citenamefont {Dagotto},\ and\ \citenamefont
  {Reboredo}}]{foyevtsovaInitioQuantumMonte2014}%
  \BibitemOpen
  \bibfield  {author} {\bibinfo {author} {\bibfnamefont {K.}~\bibnamefont
  {Foyevtsova}}, \bibinfo {author} {\bibfnamefont {J.~T.}\ \bibnamefont
  {Krogel}}, \bibinfo {author} {\bibfnamefont {J.}~\bibnamefont {Kim}},
  \bibinfo {author} {\bibfnamefont {P.~R.~C.}\ \bibnamefont {Kent}}, \bibinfo
  {author} {\bibfnamefont {E.}~\bibnamefont {Dagotto}},\ and\ \bibinfo {author}
  {\bibfnamefont {F.~A.}\ \bibnamefont {Reboredo}},\ }\bibfield  {title}
  {\bibinfo {title} {Ab initio quantum monte carlo calculations of spin
  superexchange in cuprates: The benchmarking case of
  {${\mathrm{Ca}}_{2}{\mathrm{CuO}}_{3}$}},\ }\href
  {https://doi.org/10.1103/PhysRevX.4.031003} {\bibfield  {journal} {\bibinfo
  {journal} {Physical Review X}\ }\textbf {\bibinfo {volume} {4}},\ \bibinfo
  {pages} {031003} (\bibinfo {year} {2014})}\BibitemShut {NoStop}%
\bibitem [{\citenamefont {Wagner}(2015)}]{wagnerGroundStateDoped2015}%
  \BibitemOpen
  \bibfield  {author} {\bibinfo {author} {\bibfnamefont {L.~K.}\ \bibnamefont
  {Wagner}},\ }\bibfield  {title} {\bibinfo {title} {Ground state of doped
  cuprates from first-principles quantum {{Monte Carlo}} calculations},\ }\href
  {https://doi.org/10.1103/PhysRevB.92.161116} {\bibfield  {journal} {\bibinfo
  {journal} {Physical Review B}\ }\textbf {\bibinfo {volume} {92}},\ \bibinfo
  {pages} {161116(R)} (\bibinfo {year} {2015})}\BibitemShut {NoStop}%
\bibitem [{\citenamefont {Shin}\ \emph {et~al.}(2024)\citenamefont {Shin},
  \citenamefont {Gasperich}, \citenamefont {Rojas}, \citenamefont {Ngo},
  \citenamefont {Krogel},\ and\ \citenamefont
  {Benali}}]{shinSystematicImprovementQuantum2024}%
  \BibitemOpen
  \bibfield  {author} {\bibinfo {author} {\bibfnamefont {H.}~\bibnamefont
  {Shin}}, \bibinfo {author} {\bibfnamefont {K.}~\bibnamefont {Gasperich}},
  \bibinfo {author} {\bibfnamefont {T.}~\bibnamefont {Rojas}}, \bibinfo
  {author} {\bibfnamefont {A.~T.}\ \bibnamefont {Ngo}}, \bibinfo {author}
  {\bibfnamefont {J.~T.}\ \bibnamefont {Krogel}},\ and\ \bibinfo {author}
  {\bibfnamefont {A.}~\bibnamefont {Benali}},\ }\href
  {https://doi.org/10.48550/arXiv.2403.03466} {\bibinfo {title} {Systematic
  {{Improvement}} of {{Quantum Monte Carlo Calculations}} in {{Transition Metal
  Oxides}}: {{sCI-Driven Wavefunction Optimization}} for {{Reliable Band Gap}}
  prediction}} (\bibinfo {year} {2024}),\ \Eprint
  {https://arxiv.org/abs/2403.03466} {arXiv:2403.03466 [cond-mat,
  physics:physics]} \BibitemShut {NoStop}%
\bibitem [{\citenamefont {Yu}\ \emph {et~al.}(2010)\citenamefont {Yu},
  \citenamefont {Li}, \citenamefont {Motoyama}, \citenamefont {Zhao},
  \citenamefont {Bari{\v s}i{\'c}}, \citenamefont {Cho}, \citenamefont
  {Bourges}, \citenamefont {Hradil}, \citenamefont {Mole},\ and\ \citenamefont
  {Greven}}]{yuMagneticResonanceModel2010}%
  \BibitemOpen
  \bibfield  {author} {\bibinfo {author} {\bibfnamefont {G.}~\bibnamefont
  {Yu}}, \bibinfo {author} {\bibfnamefont {Y.}~\bibnamefont {Li}}, \bibinfo
  {author} {\bibfnamefont {E.~M.}\ \bibnamefont {Motoyama}}, \bibinfo {author}
  {\bibfnamefont {X.}~\bibnamefont {Zhao}}, \bibinfo {author} {\bibfnamefont
  {N.}~\bibnamefont {Bari{\v s}i{\'c}}}, \bibinfo {author} {\bibfnamefont
  {Y.}~\bibnamefont {Cho}}, \bibinfo {author} {\bibfnamefont {P.}~\bibnamefont
  {Bourges}}, \bibinfo {author} {\bibfnamefont {K.}~\bibnamefont {Hradil}},
  \bibinfo {author} {\bibfnamefont {R.~A.}\ \bibnamefont {Mole}},\ and\
  \bibinfo {author} {\bibfnamefont {M.}~\bibnamefont {Greven}},\ }\bibfield
  {title} {\bibinfo {title} {Magnetic resonance in the model high-temperature
  superconductor {${\text{HgBa}}_{2}{\text{CuO}}_{4+\ensuremath{\delta}}$}},\
  }\href {https://doi.org/10.1103/PhysRevB.81.064518} {\bibfield  {journal}
  {\bibinfo  {journal} {Physical Review B}\ }\textbf {\bibinfo {volume} {81}},\
  \bibinfo {pages} {064518} (\bibinfo {year} {2010})}\BibitemShut {NoStop}%
\bibitem [{\citenamefont {Scalapino}(2012)}]{scalapinoCommonThreadPairing2012}%
  \BibitemOpen
  \bibfield  {author} {\bibinfo {author} {\bibfnamefont {D.~J.}\ \bibnamefont
  {Scalapino}},\ }\bibfield  {title} {\bibinfo {title} {A common thread:
  {{The}} pairing interaction for unconventional superconductors},\ }\href
  {https://doi.org/10.1103/RevModPhys.84.1383} {\bibfield  {journal} {\bibinfo
  {journal} {Reviews of Modern Physics}\ }\textbf {\bibinfo {volume} {84}},\
  \bibinfo {pages} {1383} (\bibinfo {year} {2012})}\BibitemShut {NoStop}%
\bibitem [{\citenamefont {Wagner}(2007)}]{wagnerTransitionMetalOxides2007}%
  \BibitemOpen
  \bibfield  {author} {\bibinfo {author} {\bibfnamefont {L.~K.}\ \bibnamefont
  {Wagner}},\ }\bibfield  {title} {\bibinfo {title} {Transition metal oxides
  using quantum {{Monte Carlo}}},\ }\href
  {https://doi.org/10.1088/0953-8984/19/34/343201} {\bibfield  {journal}
  {\bibinfo  {journal} {Journal of Physics: Condensed Matter}\ }\textbf
  {\bibinfo {volume} {19}},\ \bibinfo {pages} {343201} (\bibinfo {year}
  {2007})}\BibitemShut {NoStop}%
\bibitem [{\citenamefont {Sorella}\ \emph {et~al.}(2007)\citenamefont
  {Sorella}, \citenamefont {Casula},\ and\ \citenamefont
  {Rocca}}]{sorellaWeakBindingTwo2007}%
  \BibitemOpen
  \bibfield  {author} {\bibinfo {author} {\bibfnamefont {S.}~\bibnamefont
  {Sorella}}, \bibinfo {author} {\bibfnamefont {M.}~\bibnamefont {Casula}},\
  and\ \bibinfo {author} {\bibfnamefont {D.}~\bibnamefont {Rocca}},\ }\bibfield
   {title} {\bibinfo {title} {Weak binding between two aromatic rings:
  {{Feeling}} the van der {{Waals}} attraction by quantum {{Monte Carlo}}
  methods},\ }\href {https://doi.org/10.1063/1.2746035} {\bibfield  {journal}
  {\bibinfo  {journal} {The Journal of Chemical Physics}\ }\textbf {\bibinfo
  {volume} {127}},\ \bibinfo {pages} {014105} (\bibinfo {year}
  {2007})}\BibitemShut {NoStop}%
\bibitem [{\citenamefont {Di~Castro}\ \emph {et~al.}(2015)\citenamefont
  {Di~Castro}, \citenamefont {Cantoni}, \citenamefont {Ridolfi}, \citenamefont
  {Aruta}, \citenamefont {Tebano}, \citenamefont {Yang},\ and\ \citenamefont
  {Balestrino}}]{dicastroHighSuperconductivityInterface2015}%
  \BibitemOpen
  \bibfield  {author} {\bibinfo {author} {\bibfnamefont {D.}~\bibnamefont
  {Di~Castro}}, \bibinfo {author} {\bibfnamefont {C.}~\bibnamefont {Cantoni}},
  \bibinfo {author} {\bibfnamefont {F.}~\bibnamefont {Ridolfi}}, \bibinfo
  {author} {\bibfnamefont {C.}~\bibnamefont {Aruta}}, \bibinfo {author}
  {\bibfnamefont {A.}~\bibnamefont {Tebano}}, \bibinfo {author} {\bibfnamefont
  {N.}~\bibnamefont {Yang}},\ and\ \bibinfo {author} {\bibfnamefont
  {G.}~\bibnamefont {Balestrino}},\ }\bibfield  {title} {\bibinfo {title}
  {High-${T}_{c}$ superconductivity at the interface between the
  {${\mathrm{CaCuO}}_{2}$} and {${\mathrm{SrTiO}}_{3}$} insulating oxides},\
  }\href {https://doi.org/10.1103/PhysRevLett.115.147001} {\bibfield  {journal}
  {\bibinfo  {journal} {Physical Review Letters}\ }\textbf {\bibinfo {volume}
  {115}},\ \bibinfo {pages} {147001} (\bibinfo {year} {2015})}\BibitemShut
  {NoStop}%
\bibitem [{\citenamefont {Sun}\ \emph {et~al.}(2020)\citenamefont {Sun},
  \citenamefont {Zhang}, \citenamefont {Banerjee}, \citenamefont {Bao},
  \citenamefont {Barbry}, \citenamefont {Blunt}, \citenamefont {Bogdanov},
  \citenamefont {Booth}, \citenamefont {Chen}, \citenamefont {Cui},
  \citenamefont {Eriksen}, \citenamefont {Gao}, \citenamefont {Guo},
  \citenamefont {Hermann}, \citenamefont {Hermes}, \citenamefont {Koh},
  \citenamefont {Koval}, \citenamefont {Lehtola}, \citenamefont {Li},
  \citenamefont {Liu}, \citenamefont {Mardirossian}, \citenamefont {McClain},
  \citenamefont {Motta}, \citenamefont {Mussard}, \citenamefont {Pham},
  \citenamefont {Pulkin}, \citenamefont {Purwanto}, \citenamefont {Robinson},
  \citenamefont {Ronca}, \citenamefont {Sayfutyarova}, \citenamefont
  {Scheurer}, \citenamefont {Schurkus}, \citenamefont {Smith}, \citenamefont
  {Sun}, \citenamefont {Sun}, \citenamefont {Upadhyay}, \citenamefont {Wagner},
  \citenamefont {Wang}, \citenamefont {White}, \citenamefont {Whitfield},
  \citenamefont {Williamson}, \citenamefont {Wouters}, \citenamefont {Yang},
  \citenamefont {Yu}, \citenamefont {Zhu}, \citenamefont {Berkelbach},
  \citenamefont {Sharma}, \citenamefont {Sokolov},\ and\ \citenamefont
  {Chan}}]{sunRecentDevelopmentsPySCF2020}%
  \BibitemOpen
  \bibfield  {author} {\bibinfo {author} {\bibfnamefont {Q.}~\bibnamefont
  {Sun}}, \bibinfo {author} {\bibfnamefont {X.}~\bibnamefont {Zhang}}, \bibinfo
  {author} {\bibfnamefont {S.}~\bibnamefont {Banerjee}}, \bibinfo {author}
  {\bibfnamefont {P.}~\bibnamefont {Bao}}, \bibinfo {author} {\bibfnamefont
  {M.}~\bibnamefont {Barbry}}, \bibinfo {author} {\bibfnamefont {N.~S.}\
  \bibnamefont {Blunt}}, \bibinfo {author} {\bibfnamefont {N.~A.}\ \bibnamefont
  {Bogdanov}}, \bibinfo {author} {\bibfnamefont {G.~H.}\ \bibnamefont {Booth}},
  \bibinfo {author} {\bibfnamefont {J.}~\bibnamefont {Chen}}, \bibinfo {author}
  {\bibfnamefont {Z.-H.}\ \bibnamefont {Cui}}, \bibinfo {author} {\bibfnamefont
  {J.~J.}\ \bibnamefont {Eriksen}}, \bibinfo {author} {\bibfnamefont
  {Y.}~\bibnamefont {Gao}}, \bibinfo {author} {\bibfnamefont {S.}~\bibnamefont
  {Guo}}, \bibinfo {author} {\bibfnamefont {J.}~\bibnamefont {Hermann}},
  \bibinfo {author} {\bibfnamefont {M.~R.}\ \bibnamefont {Hermes}}, \bibinfo
  {author} {\bibfnamefont {K.}~\bibnamefont {Koh}}, \bibinfo {author}
  {\bibfnamefont {P.}~\bibnamefont {Koval}}, \bibinfo {author} {\bibfnamefont
  {S.}~\bibnamefont {Lehtola}}, \bibinfo {author} {\bibfnamefont
  {Z.}~\bibnamefont {Li}}, \bibinfo {author} {\bibfnamefont {J.}~\bibnamefont
  {Liu}}, \bibinfo {author} {\bibfnamefont {N.}~\bibnamefont {Mardirossian}},
  \bibinfo {author} {\bibfnamefont {J.~D.}\ \bibnamefont {McClain}}, \bibinfo
  {author} {\bibfnamefont {M.}~\bibnamefont {Motta}}, \bibinfo {author}
  {\bibfnamefont {B.}~\bibnamefont {Mussard}}, \bibinfo {author} {\bibfnamefont
  {H.~Q.}\ \bibnamefont {Pham}}, \bibinfo {author} {\bibfnamefont
  {A.}~\bibnamefont {Pulkin}}, \bibinfo {author} {\bibfnamefont
  {W.}~\bibnamefont {Purwanto}}, \bibinfo {author} {\bibfnamefont {P.~J.}\
  \bibnamefont {Robinson}}, \bibinfo {author} {\bibfnamefont {E.}~\bibnamefont
  {Ronca}}, \bibinfo {author} {\bibfnamefont {E.~R.}\ \bibnamefont
  {Sayfutyarova}}, \bibinfo {author} {\bibfnamefont {M.}~\bibnamefont
  {Scheurer}}, \bibinfo {author} {\bibfnamefont {H.~F.}\ \bibnamefont
  {Schurkus}}, \bibinfo {author} {\bibfnamefont {J.~E.~T.}\ \bibnamefont
  {Smith}}, \bibinfo {author} {\bibfnamefont {C.}~\bibnamefont {Sun}}, \bibinfo
  {author} {\bibfnamefont {S.-N.}\ \bibnamefont {Sun}}, \bibinfo {author}
  {\bibfnamefont {S.}~\bibnamefont {Upadhyay}}, \bibinfo {author}
  {\bibfnamefont {L.~K.}\ \bibnamefont {Wagner}}, \bibinfo {author}
  {\bibfnamefont {X.}~\bibnamefont {Wang}}, \bibinfo {author} {\bibfnamefont
  {A.}~\bibnamefont {White}}, \bibinfo {author} {\bibfnamefont {J.~D.}\
  \bibnamefont {Whitfield}}, \bibinfo {author} {\bibfnamefont {M.~J.}\
  \bibnamefont {Williamson}}, \bibinfo {author} {\bibfnamefont
  {S.}~\bibnamefont {Wouters}}, \bibinfo {author} {\bibfnamefont
  {J.}~\bibnamefont {Yang}}, \bibinfo {author} {\bibfnamefont {J.~M.}\
  \bibnamefont {Yu}}, \bibinfo {author} {\bibfnamefont {T.}~\bibnamefont
  {Zhu}}, \bibinfo {author} {\bibfnamefont {T.~C.}\ \bibnamefont {Berkelbach}},
  \bibinfo {author} {\bibfnamefont {S.}~\bibnamefont {Sharma}}, \bibinfo
  {author} {\bibfnamefont {A.~Y.}\ \bibnamefont {Sokolov}},\ and\ \bibinfo
  {author} {\bibfnamefont {G.~K.-L.}\ \bibnamefont {Chan}},\ }\bibfield
  {title} {\bibinfo {title} {Recent developments in the {{PySCF}} program
  package},\ }\href {https://doi.org/10.1063/5.0006074} {\bibfield  {journal}
  {\bibinfo  {journal} {The Journal of Chemical Physics}\ }\textbf {\bibinfo
  {volume} {153}},\ \bibinfo {pages} {024109} (\bibinfo {year}
  {2020})}\BibitemShut {NoStop}%
\bibitem [{\citenamefont {Sun}\ \emph {et~al.}(2018)\citenamefont {Sun},
  \citenamefont {Berkelbach}, \citenamefont {Blunt}, \citenamefont {Booth},
  \citenamefont {Guo}, \citenamefont {Li}, \citenamefont {Liu}, \citenamefont
  {McClain}, \citenamefont {Sayfutyarova}, \citenamefont {Sharma},
  \citenamefont {Wouters},\ and\ \citenamefont
  {Chan}}]{sunPySCFPythonbasedSimulations2018}%
  \BibitemOpen
  \bibfield  {author} {\bibinfo {author} {\bibfnamefont {Q.}~\bibnamefont
  {Sun}}, \bibinfo {author} {\bibfnamefont {T.~C.}\ \bibnamefont {Berkelbach}},
  \bibinfo {author} {\bibfnamefont {N.~S.}\ \bibnamefont {Blunt}}, \bibinfo
  {author} {\bibfnamefont {G.~H.}\ \bibnamefont {Booth}}, \bibinfo {author}
  {\bibfnamefont {S.}~\bibnamefont {Guo}}, \bibinfo {author} {\bibfnamefont
  {Z.}~\bibnamefont {Li}}, \bibinfo {author} {\bibfnamefont {J.}~\bibnamefont
  {Liu}}, \bibinfo {author} {\bibfnamefont {J.~D.}\ \bibnamefont {McClain}},
  \bibinfo {author} {\bibfnamefont {E.~R.}\ \bibnamefont {Sayfutyarova}},
  \bibinfo {author} {\bibfnamefont {S.}~\bibnamefont {Sharma}}, \bibinfo
  {author} {\bibfnamefont {S.}~\bibnamefont {Wouters}},\ and\ \bibinfo {author}
  {\bibfnamefont {G.~K.-L.}\ \bibnamefont {Chan}},\ }\bibfield  {title}
  {\bibinfo {title} {{{PySCF}}: The {{Python-based}} simulations of chemistry
  framework},\ }\href {https://doi.org/10.1002/wcms.1340} {\bibfield  {journal}
  {\bibinfo  {journal} {WIREs Computational Molecular Science}\ }\textbf
  {\bibinfo {volume} {8}},\ \bibinfo {pages} {e1340} (\bibinfo {year}
  {2018})}\BibitemShut {NoStop}%
\bibitem [{\citenamefont {Sun}(2015)}]{sunLibcintEfficientGeneral2015}%
  \BibitemOpen
  \bibfield  {author} {\bibinfo {author} {\bibfnamefont {Q.}~\bibnamefont
  {Sun}},\ }\bibfield  {title} {\bibinfo {title} {Libcint: {{An}} efficient
  general integral library for {{Gaussian}} basis functions},\ }\href
  {https://doi.org/10.1002/jcc.23981} {\bibfield  {journal} {\bibinfo
  {journal} {Journal of Computational Chemistry}\ }\textbf {\bibinfo {volume}
  {36}},\ \bibinfo {pages} {1664} (\bibinfo {year} {2015})}\BibitemShut
  {NoStop}%
\bibitem [{\citenamefont {Bennett}\ \emph {et~al.}(2017)\citenamefont
  {Bennett}, \citenamefont {Melton}, \citenamefont {Annaberdiyev},
  \citenamefont {Wang}, \citenamefont {Shulenburger},\ and\ \citenamefont
  {Mitas}}]{bennettNewGenerationEffective2017}%
  \BibitemOpen
  \bibfield  {author} {\bibinfo {author} {\bibfnamefont {M.~C.}\ \bibnamefont
  {Bennett}}, \bibinfo {author} {\bibfnamefont {C.~A.}\ \bibnamefont {Melton}},
  \bibinfo {author} {\bibfnamefont {A.}~\bibnamefont {Annaberdiyev}}, \bibinfo
  {author} {\bibfnamefont {G.}~\bibnamefont {Wang}}, \bibinfo {author}
  {\bibfnamefont {L.}~\bibnamefont {Shulenburger}},\ and\ \bibinfo {author}
  {\bibfnamefont {L.}~\bibnamefont {Mitas}},\ }\bibfield  {title} {\bibinfo
  {title} {A new generation of effective core potentials for correlated
  calculations},\ }\href {https://doi.org/10.1063/1.4995643} {\bibfield
  {journal} {\bibinfo  {journal} {The Journal of Chemical Physics}\ }\textbf
  {\bibinfo {volume} {147}},\ \bibinfo {pages} {224106} (\bibinfo {year}
  {2017})}\BibitemShut {NoStop}%
\bibitem [{\citenamefont {Annaberdiyev}\ \emph {et~al.}(2018)\citenamefont
  {Annaberdiyev}, \citenamefont {Wang}, \citenamefont {Melton}, \citenamefont
  {Bennett}, \citenamefont {Shulenburger},\ and\ \citenamefont
  {Mitas}}]{annaberdiyevNewGenerationEffective2018}%
  \BibitemOpen
  \bibfield  {author} {\bibinfo {author} {\bibfnamefont {A.}~\bibnamefont
  {Annaberdiyev}}, \bibinfo {author} {\bibfnamefont {G.}~\bibnamefont {Wang}},
  \bibinfo {author} {\bibfnamefont {C.~A.}\ \bibnamefont {Melton}}, \bibinfo
  {author} {\bibfnamefont {M.~C.}\ \bibnamefont {Bennett}}, \bibinfo {author}
  {\bibfnamefont {L.}~\bibnamefont {Shulenburger}},\ and\ \bibinfo {author}
  {\bibfnamefont {L.}~\bibnamefont {Mitas}},\ }\bibfield  {title} {\bibinfo
  {title} {A new generation of effective core potentials from correlated
  calculations: 3d transition metal series},\ }\href
  {https://doi.org/10.1063/1.5040472} {\bibfield  {journal} {\bibinfo
  {journal} {The Journal of Chemical Physics}\ }\textbf {\bibinfo {volume}
  {149}},\ \bibinfo {pages} {134108} (\bibinfo {year} {2018})}\BibitemShut
  {NoStop}%
\bibitem [{\citenamefont {Wang}\ \emph {et~al.}(2019)\citenamefont {Wang},
  \citenamefont {Annaberdiyev}, \citenamefont {Melton}, \citenamefont
  {Bennett}, \citenamefont {Shulenburger},\ and\ \citenamefont
  {Mitas}}]{wangNewGenerationEffective2019}%
  \BibitemOpen
  \bibfield  {author} {\bibinfo {author} {\bibfnamefont {G.}~\bibnamefont
  {Wang}}, \bibinfo {author} {\bibfnamefont {A.}~\bibnamefont {Annaberdiyev}},
  \bibinfo {author} {\bibfnamefont {C.~A.}\ \bibnamefont {Melton}}, \bibinfo
  {author} {\bibfnamefont {M.~C.}\ \bibnamefont {Bennett}}, \bibinfo {author}
  {\bibfnamefont {L.}~\bibnamefont {Shulenburger}},\ and\ \bibinfo {author}
  {\bibfnamefont {L.}~\bibnamefont {Mitas}},\ }\bibfield  {title} {\bibinfo
  {title} {A new generation of effective core potentials from correlated
  calculations: 4s and 4p main group elements and first row additions},\ }\href
  {https://doi.org/10.1063/1.5121006} {\bibfield  {journal} {\bibinfo
  {journal} {The Journal of Chemical Physics}\ }\textbf {\bibinfo {volume}
  {151}},\ \bibinfo {pages} {144110} (\bibinfo {year} {2019})}\BibitemShut
  {NoStop}%
\bibitem [{\citenamefont {Sayfutyarova}\ \emph {et~al.}(2017)\citenamefont
  {Sayfutyarova}, \citenamefont {Sun}, \citenamefont {Chan},\ and\
  \citenamefont {Knizia}}]{sayfutyarovaAutomatedConstructionMolecular2017}%
  \BibitemOpen
  \bibfield  {author} {\bibinfo {author} {\bibfnamefont {E.~R.}\ \bibnamefont
  {Sayfutyarova}}, \bibinfo {author} {\bibfnamefont {Q.}~\bibnamefont {Sun}},
  \bibinfo {author} {\bibfnamefont {G.~K.-L.}\ \bibnamefont {Chan}},\ and\
  \bibinfo {author} {\bibfnamefont {G.}~\bibnamefont {Knizia}},\ }\bibfield
  {title} {\bibinfo {title} {Automated {{Construction}} of {{Molecular Active
  Spaces}} from {{Atomic Valence Orbitals}}},\ }\href
  {https://doi.org/10.1021/acs.jctc.7b00128} {\bibfield  {journal} {\bibinfo
  {journal} {Journal of Chemical Theory and Computation}\ }\textbf {\bibinfo
  {volume} {13}},\ \bibinfo {pages} {4063} (\bibinfo {year}
  {2017})}\BibitemShut {NoStop}%
\bibitem [{\citenamefont {Wheeler}\ \emph {et~al.}(2023)\citenamefont
  {Wheeler}, \citenamefont {Pathak}, \citenamefont {Kleiner}, \citenamefont
  {Yuan}, \citenamefont {Rodrigues}, \citenamefont {Lorsung}, \citenamefont
  {Krongchon}, \citenamefont {Chang}, \citenamefont {Zhou}, \citenamefont
  {Busemeyer}, \citenamefont {Williams}, \citenamefont {Mu{\~n}oz},
  \citenamefont {Chow},\ and\ \citenamefont
  {Wagner}}]{wheelerPyQMCAllPythonRealspace2023}%
  \BibitemOpen
  \bibfield  {author} {\bibinfo {author} {\bibfnamefont {W.~A.}\ \bibnamefont
  {Wheeler}}, \bibinfo {author} {\bibfnamefont {S.}~\bibnamefont {Pathak}},
  \bibinfo {author} {\bibfnamefont {K.~G.}\ \bibnamefont {Kleiner}}, \bibinfo
  {author} {\bibfnamefont {S.}~\bibnamefont {Yuan}}, \bibinfo {author}
  {\bibfnamefont {J.~N.~B.}\ \bibnamefont {Rodrigues}}, \bibinfo {author}
  {\bibfnamefont {C.}~\bibnamefont {Lorsung}}, \bibinfo {author} {\bibfnamefont
  {K.}~\bibnamefont {Krongchon}}, \bibinfo {author} {\bibfnamefont
  {Y.}~\bibnamefont {Chang}}, \bibinfo {author} {\bibfnamefont
  {Y.}~\bibnamefont {Zhou}}, \bibinfo {author} {\bibfnamefont {B.}~\bibnamefont
  {Busemeyer}}, \bibinfo {author} {\bibfnamefont {K.~T.}\ \bibnamefont
  {Williams}}, \bibinfo {author} {\bibfnamefont {A.}~\bibnamefont {Mu{\~n}oz}},
  \bibinfo {author} {\bibfnamefont {C.~Y.}\ \bibnamefont {Chow}},\ and\
  \bibinfo {author} {\bibfnamefont {L.~K.}\ \bibnamefont {Wagner}},\ }\bibfield
   {title} {\bibinfo {title} {{{PyQMC}}: {{An}} all-{{Python}} real-space
  quantum {{Monte Carlo}} module in {{PySCF}}},\ }\href
  {https://doi.org/10.1063/5.0139024} {\bibfield  {journal} {\bibinfo
  {journal} {The Journal of Chemical Physics}\ }\textbf {\bibinfo {volume}
  {158}},\ \bibinfo {pages} {114801} (\bibinfo {year} {2023})}\BibitemShut
  {NoStop}%
\bibitem [{\citenamefont {Wagner}\ and\ \citenamefont
  {Mitas}(2007)}]{wagnerEnergeticsDipoleMoment2007}%
  \BibitemOpen
  \bibfield  {author} {\bibinfo {author} {\bibfnamefont {L.~K.}\ \bibnamefont
  {Wagner}}\ and\ \bibinfo {author} {\bibfnamefont {L.}~\bibnamefont {Mitas}},\
  }\bibfield  {title} {\bibinfo {title} {Energetics and dipole moment of
  transition metal monoxides by quantum {{Monte Carlo}}},\ }\href
  {https://doi.org/10.1063/1.2428294} {\bibfield  {journal} {\bibinfo
  {journal} {The Journal of Chemical Physics}\ }\textbf {\bibinfo {volume}
  {126}},\ \bibinfo {pages} {034105} (\bibinfo {year} {2007})}\BibitemShut
  {NoStop}%
\bibitem [{\citenamefont
  {Casula}(2006)}]{casulaLocalityApproximationStandard2006}%
  \BibitemOpen
  \bibfield  {author} {\bibinfo {author} {\bibfnamefont {M.}~\bibnamefont
  {Casula}},\ }\bibfield  {title} {\bibinfo {title} {Beyond the locality
  approximation in the standard diffusion {{Monte Carlo}} method},\ }\href
  {https://doi.org/10.1103/PhysRevB.74.161102} {\bibfield  {journal} {\bibinfo
  {journal} {Physical Review B}\ }\textbf {\bibinfo {volume} {74}},\ \bibinfo
  {pages} {161102(R)} (\bibinfo {year} {2006})}\BibitemShut {NoStop}%
\bibitem [{\citenamefont {Anderson}\ and\ \citenamefont
  {Umrigar}(2021)}]{andersonNonlocalPseudopotentialsTimestep2021}%
  \BibitemOpen
  \bibfield  {author} {\bibinfo {author} {\bibfnamefont {T.~A.}\ \bibnamefont
  {Anderson}}\ and\ \bibinfo {author} {\bibfnamefont {C.~J.}\ \bibnamefont
  {Umrigar}},\ }\bibfield  {title} {\bibinfo {title} {Nonlocal pseudopotentials
  and time-step errors in diffusion {{Monte Carlo}}},\ }\href
  {https://doi.org/10.1063/5.0052838} {\bibfield  {journal} {\bibinfo
  {journal} {The Journal of Chemical Physics}\ }\textbf {\bibinfo {volume}
  {154}},\ \bibinfo {pages} {214110} (\bibinfo {year} {2021})}\BibitemShut
  {NoStop}%
\end{thebibliography}%

\end{document}